\documentclass[a4paper,11pt]{article}
\pdfoutput=1 

\usepackage{jheppub, amsmath, amssymb,amsthm,dsfont,graphicx,mathtools,multirow,rotating,placeins,verbatim} 

\usepackage{tensor}
\usepackage[T1]{fontenc} 
\usepackage[all,color]{xy}

\newcommand{\sst}[1]{{\scriptscriptstyle #1}}

\newcommand{\dbar}{{d\mkern-7mu\mathchar'26\mkern-2mu}}

\def\0{{\sst{(0)}}}
\def\1{{\sst{(1)}}}
\def\2{{\sst{(2)}}}
\def\3{{\sst{(3)}}}
\def\4{{\sst{(4)}}}
\def\5{{\sst{(5)}}}
\def\6{{\sst{(6)}}}
\def\7{{\sst{(7)}}}

\newcommand{\be}{\begin{equation}}
\newcommand{\ee}{\end{equation}}
\def\ba{\begin{array}}
\def\ea{\end{array}}

\newcommand{\bea}{\begin{eqnarray}}
\newcommand{\eea}{\end{eqnarray}}

\DeclareMathOperator{\tr}{tr}

\newcommand{\N}{\mathcal{N}}

\title{\boldmath The pure BRST Einstein-Hilbert Lagrangian from the double-copy to cubic order}

\author[a, b]{L. Borsten,}
\author[c, 1]{S. Nagy\note{Corresponding author.}}

\affiliation[a]{Maxwell Institute for Mathematical Sciences,\\ 
Department of Mathematics, Heriot-Watt University,\\
 Colin Maclaurin Building, Riccarton, Edinburgh EH14 4AS, United Kingdom}

\affiliation[b]{School of Theoretical Physics, Dublin Institute for Advanced Studies,\\
10 Burlington Road, Dublin 4, Ireland}
\affiliation[c]{Centre for Astronomy \& Particle Theory,
University Park,
Nottingham,
NG7 2RD,
United Kingdom}

\emailAdd{l.borsten@hw.ac.uk}
\emailAdd{silvia.nagy@nottingham.ac.uk}

\abstract{ We construct  the pure gravity  Becchi-Rouet-Stora-Tyutin (BRST) Einstein-Hilbert Lagrangian, to cubic order, using the BRST convolution product of two Yang-Mills theories, in conjunction with the Bern-Carrasco-Johansson (BCJ) double-copy.}

\begin{document} 
\maketitle
\flushbottom

\section{Introduction}

 In the present contribution we construct  the pure gravity Lagrangian, to cubic order, using the BRST convolution product of two Yang-Mills theories introduced in \cite{Anastasiou:2014qba,Anastasiou:2018rdx}, in conjunction with the BCJ double-copy \cite{Bern:2008qj, Bern:2010ue,  Bern:2010yg} and, in particular, the developments of  \cite{Bern:2010yg, Luna:2016hge}.

 The BCJ colour-kinematic duality conjecture \cite{Bern:2008qj, Bern:2010ue,  Bern:2010yg} implies that the amplitudes of $\N=0$ supergravity,  
\be
S_{\N=0} =\frac{1}{2\kappa^2} \int \star  R -\frac{1}{(D-2)}  d\varphi \wedge \star  d\varphi - \frac{1}{2}e^{-\frac{4}{D-2}\varphi}  H \wedge  \star  H,
\ee
where  $\varphi$  is the dilaton and  $H=dB$ is  the Kalb-Ramond  (KR) 2-form   field strength, follow from the double-copy of the amplitudes of pure Yang-Mills theory with arbitrary non-Abelian gauge group,
\be
S_{\text{YM}} = \frac{1}{2g^2} \tr \int F\wedge \star F,
\ee
to all orders in perturbation theory.

Recall,   $\N=0$ supergravity is the common NS-NS sector of the $\alpha'\rightarrow 0$ limit of closed string theories. In this context, its appearance is understood to be a consequence of ``open $\times$ open $=$ closed'' property of the string spectrum underlying  the Kawai-Lewellen-Tye (KLT) tree-level scattering amplitude relations of  string theory \cite{Kawai:1985xq}. At the level of the massless on-shell states we have the straighforward tensor product,
\be\label{states}
A_i \otimes A_j = h_{ij} \oplus B_{ij} \oplus \varphi,
\ee
where $i, j=1,\ldots, D-2$.

The   BCJ duality for gluons has been established to all orders at tree-level from a number of perspectives \cite{Kiermaier:2010, BjerrumBohr:2010hn, Mafra:2011kj, Du:2016tbc} and has been generalised to include numerous (super) Yang-Mills theories  \cite{Bern:2009kd, Bern:2010ue, Bern:2010yg, Chiodaroli:2011pp, Bern:2012gh, Carrasco:2012ca, Damgaard:2012fb, Huang:2012wr, Bargheer:2012gv, Carrasco:2013ypa,  Chiodaroli:2013upa, Johansson:2014zca, Chiodaroli:2014xia, Chiodaroli:2015rdg, Chiodaroli:2015wal, Chiodaroli:2016jqw, Carrasco:2016ygv, Carrasco:2016ldy, Anastasiou:2016csv, Johansson:2017bfl, Johansson:2017srf, Azevedo:2017lkz, Anastasiou:2017nsz, Chiodaroli:2017ngp, Chiodaroli:2017ehv,  Chiodaroli:2018dbu,  Bern:2017ucb, Azevedo:2018dgo, Momeni:2020vvr}, generating a wide variety of double-copy constructible  gravity theories. Although BCJ duality remains conjectural at loop-level, there is a growing list of highly non-trivial examples \cite{Bern:2010ue,Bern:2010tq,Carrasco:2011mn,Bern:2011rj,BoucherVeronneau:2011qv,Bern:2012cd,Bern:2012gh, Oxburgh:2012zr,Bern:2012uf,Du:2012mt,Yuan:2012rg,Bern:2013uka, Boels:2013bi,Bern:2013yya,Bern:2013qca, Bern:2014lha, Bern:2014sna, Mafra:2015mja, Johansson:2017bfl,  Bern:2017ucb, Bern:2018jmv}.

This programme is suggestive of a deep ``gravity $=$ gauge $\times$ gauge'' relation and has already dramatically advanced our understanding of perturbative  quantum gravity. See for example \cite{Bern:2009kd,Bern:2012cd, Bern:2014lha,Bern:2014sna, Bern:2015xsa, Bern:2018jmv}.   This motivates an effort to understand the degree to which the paradigm can be pushed beyond  scattering amplitudes and the identification of asymptotic on-shell states as in \eqref{states}. For example, it has been shown that one can manifest  BCJ duality  at the level of the Lagrangian or field equations from a number of points of view \cite{Bern:2008qj, Bern:2010ue, Monteiro:2011pc, BjerrumBohr:2012mg, Tolotti:2013caa, Monteiro:2013rya, Fu:2016plh,  Cheung:2016prv, Chen:2019ywi}.  A related line of research has been the construction of classical solutions in theories of gravity, such as black holes,  from gauge theory solutions. This comes either in the form of applying a classical  double-copy-type map to classical gauge theory solutions or extracting perturbative classical solutions from the double-copy of gauge theory  amplitudes  \cite{Monteiro:2014cda, Luna:2015paa,Ridgway:2015fdl,  Luna:2016due, White:2016jzc,Goldberger:2016iau,  Cardoso:2016ngt, Cardoso:2016amd,Luna:2020adi, Luna:2016hge, Goldberger:2017frp, LopesCardoso:2018xes, Luna:2017dtq, Bahjat-Abbas:2017htu, Berman:2018hwd, Plefka:2018dpa, Bahjat-Abbas:2018vgo,Luna:2018dpt,Shen:2018ebu,  Cheung:2018wkq, Kosower:2018adc, CarrilloGonzalez:2019gof, Johansson:2019dnu,   Maybee:2019jus,Plefka:2019hmz,  Bern:2019nnu,  Bern:2019crd, Arkani-Hamed:2019ymq, Alawadhi:2019urr, Bah:2019sda, Plefka:2019wyg,Banerjee:2019saj, Alfonsi:2020lub,Kim:2019jwm,Bahjat-Abbas:2020cyb, Lescano:2020nve, Bern:2020gjj}. This has yielded both pragmatic applications,  particularly in the context of gravity-wave astronomy,  as well as emphasising interesting questions and features of the double-copy itself. For a  review of these ideas, including the many topics not mentioned here, their applications and further references see \cite{Carrasco:2015iwa, Bern:2019prr,Borsten:2020bgv}.  

The approach taken here makes key use of  the  field theoretic convolutive product of gauge theories, introduced  in \cite{Anastasiou:2014qba} and further developed in \cite{Borsten:2013bp,  Anastasiou:2016csv, Cardoso:2016ngt, Cardoso:2016amd, Borsten:2017jpt, LopesCardoso:2018xes, Anastasiou:2018rdx, Luna:2020adi}. Using the product, the local   symmetries and equations of motion of the resulting gravity theory have been shown to arise from those of the gauge theory factors, to linear order, making crucial use of the the BRST formalism  \cite{Borsten:2017jpt,  Anastasiou:2018rdx, Luna:2020adi, Borsten:2019prq}. The procedure was recently extended beyond Minkowski backgrounds in \cite{Borsten:2019prq} and applied to the Janis-Newman-Winicour (JNW) solution in \cite{Luna:2020adi}.
 Given enough global symmetries in the gauge theory factors, this can then be used to identify the corresponding gravitational theory and all its symmetries uniquely  \cite{Borsten:2013bp, Anastasiou:2013hba, Anastasiou:2014qba, Nagy:2014jza,  Borsten:2015pla, Anastasiou:2015vba, Anastasiou:2016csv, Cardoso:2016ngt, Cardoso:2016amd, Anastasiou:2017nsz, Borsten:2017jpt,Borsten:2018jjm}.  The field theoretic product is \emph{a priori} independent from the BCJ procedure, however it is consistent with it in the sense that the double-copy amplitudes correspond to the theory obtained from the field product, as seen from the matching of symmetries. 
 
Another  important and early development \cite{Bern:2010yg} particularly relevant here, promotes BCJ duality and the double-copy to the level of actions.   First, it was shown  that  the Yang-Mills actions may be put into  a purely cubic form that manifestly yields   colour-kinematic dual tree-level amplitudes by introducing an infinite tower of auxiliary fields \cite{Bern:2010yg}. This was demonstrated explicitly to five-points in \cite{Bern:2010yg} and a systematic formulation of the manifestly tree-level BCJ respecting action to all orders was given in \cite{Tolotti:2013caa}, albeit without auxiliary fields so that it is necessarily non-local. Given Yang-Mills theory written in such a  form, the double-copy principle can be straightforwardly applied to generate  an action that correctly reproduces all the tree-level amplitudes of perturbative $\N=0$ supergravity, as described in \cite{Bern:2010yg} and developed in context of perturbative solutions in \cite{Luna:2016hge}.

One of the main advantages of the BRST formulation of the field theory convolution product  is that it gives an elegant solution to the issue of the mixing of the dilaton and graviton degrees of freedom pointed out in e.g.~\cite{LopesCardoso:2018xes,Luna:2016hge}, as detailed in \cite{Anastasiou:2018rdx,Borsten:2019prq,Borsten:2020bgv,Luna:2020adi}\footnote{For related considerations in the context of the classical Kerr-Schild double-copy see \cite{Kim:2019jwm}.}.  In particular, the BRST approach opens the door to an off-shell construction for pure Einstein-Hilbert gravity. As such, it is desirable to extend the construction beyond linear order. To this end we employ the field theory convolution product and BRST prescription of \cite{Anastasiou:2018rdx,Borsten:2019prq} in tandem with the  double-copy applied to actions, as developed in \cite{Bern:2010yg, Luna:2016hge}, to reproduce the BRST Einstein-Hilbert Lagrangian to cubic order. 

The paper is organised as follows: we give an overview of the linearised BRST double-copy procedure in \autoref{A linearised tale told again (with ghosts)}, applied to the simplest case of the product of two gauge fields. We demonstrate in particular how the ghost fields allow us to truncate to pure gravity in a robust way. We extend the construction to cubic order in perturbation theory in \autoref{Einstein-Hilbert from the double-copy to cubic order}, demonstrating that the BCJ construction applied to the BRST Yang-Mills action to cubic order reproduces Einstein-Hilbert gravity, up to field redefinitions. We also give an algorithm for mapping between the gauge-fixing functionals of the gauge theory and gravity sides. We conclude in  \autoref{Conclusions}.

\section{A linearised tale told again (with ghosts)}\label{A linearised tale told again (with ghosts)}

 \subsection{Review of convolution dictionary and the necessity of the BRST framework} 
At linear level, the double copy dictionary is constructed from an associative  convolutive inner tensor product with respect to the Poincar\'e group
\be
\label{flat_convo_def}
[f\cdot g](x)=\int d^Dy f(y) \otimes g(x-y).
\ee
We will use the notation
\begin{equation}\label{product}
[f \circ   \tilde{f}](x) = [f^a \cdot \Phi_{a\tilde{a}} \cdot \tilde{f}^{\tilde{a}}](x),
\end{equation}
 where $ \Phi$ is the convolutive pseudo-inverse $\Phi = \phi^{-1}$, with $\phi\cdot\Phi\cdot\phi=\phi$ of the  bi-adjoint scalar $\phi$ of the BCJ zeroth-copy \cite{Hodges:2011wm, Vaman:2010ez, Cachazo:2013iea, Monteiro:2013rya, Cachazo:2014xea,Monteiro:2014cda, Chiodaroli:2014xia,  Naculich:2014naa, Luna:2015paa, Naculich:2015zha, Chiodaroli:2015rdg, Luna:2016due, White:2016jzc, DeSmet:2017rve, Cheung:2016prv, Chiodaroli:2017ngp, Brown:2018wss}. Note the circle product can be generalised to include fundamental matter fields, by including a bi-fundamental scalar field \cite{Anastasiou:2016csv}.   The product \eqref{product} applied to  left $A_\mu$ and right $\tilde{A}_\nu$ pure Yang-Mills theories would be expected to yield $\N=0$ supergravity off-shell, given the BCJ amplitude relations and the tensor product of the on-shell states \eqref{states}.

However, this expectation is only met once the BRST formalism is incorporated. This can be traced back to a number of issues that have been identified in relation to this construction in the context of off-shell or classical approaches:
\begin{itemize}
\item It is  difficult to disentangle the graviton and dilaton degrees of freedom \cite{LopesCardoso:2018xes,Luna:2017dtq}. A formal demonstration of this is presented in \cite{LopesCardoso:2018xes}. Let  $j_\mu$ and $\tilde{j}_\mu$ be the sources of the   Yang-Mills  equation of motion, $j_{\mu\nu}^{(h)}$ the graviton source and $j^{(\varphi)}$ the dilaton source. Then we have 
\be 
\label{source_issue}
j^{(\varphi)}\propto j_{\ \rho}^{(h) \rho}\propto \tfrac{1}{\square}j_\rho\circ \tilde{j}^\rho \ .
\ee
Thus we see that the graviton and dilaton sources are not independent. In particular, choosing to not source the dilaton will severely restrict the graviton. We can interpret this as a constraint on gravitational theories that  admit a double copy description, appearing already at the linear order. 
\item The above comments are  a general feature of the classical BCJ double-copy, and not a consequence of the set-up  in  \cite{LopesCardoso:2018xes}.   This is evident in the  mismatch between the on-shell and off-shell degrees of freedom. Specifically,   $A_\mu\times\tilde{A}_\nu$ has $3 \times 3$ degrees of freedom off-shell, which is  insufficient  to describe the ten off-shell degrees of freedom carried by the graviton--two-form--dilaton system \cite{LopesCardoso:2018xes,Anastasiou:2018rdx}. This issue only becomes more apparent  with the addition of supersymmetry, where one lacks a full supermultiplets of off-shell degrees of freedom \cite{Anastasiou:2014qba,Siegel:1988qu,Siegel:1995px}.   


\item The  classical double-copy is usually formulated with some specific gauge fixing on both the  Yang-Mills  and the gravity side. However, there is no general procedure determining a mapping between these corresponding gauge choices, potentially introducing ambiguities into the  double-copy when taken beyond the domain of amplitudes.
\end{itemize} 
The BRST dictionary in \cite{Anastasiou:2018rdx}  resolves the above issues by taking products of sets of fields $(A_\mu,c^\alpha)$ and $(\tilde{A}_\mu,\tilde{c}^\alpha)$. Here $c^1=c$ and $c^2=\bar{c}$ are the Fadeev-Popov ghost and antighost, respectively. The off-shell d.o.f.~of the  $(A_\mu,c^\alpha)\times(\tilde{A}_\mu,\tilde{c}^\alpha)$ product can now be seen to correspond to those of the linearised BRST systems for the graviton, two-form and dilaton\footnote{Note that the d.o.f. counting is now graded by ghost number - see  \cite{Anastasiou:2018rdx} for details.}. It also naturally incorporates the ghost and ghost-for-ghost transformations \cite{Anastasiou:2018rdx,BorstenXXX}. 

We will describe in \autoref{Killing the dilaton and two-form to get pure gravity} how the BRST procedure resolves the source issue \eqref{source_issue}, and allows us to obtain a pure gravity theory. We will also present the gauge mapping algorithm between between pure Yang-Mills  theory and gravity coupled to a KR 2-form and a dilaton in \autoref{Dictionary and gauge mapping}.

\subsection{Dictionary and gauge mapping}\label{Dictionary and gauge mapping}
The general form of the BRST action is (having eliminated the Nakanishi-Lautrup auxiliary field), schematically:
\be 
S_{\text{BRST}}=\int d^D x\left(\mathcal{L}_0[f]+\tfrac{1}{2\xi}G[f]^2-\bar{c}Q\left(G[f]\right) \right)-f j^{(f)}+\bar{j}c+\bar{c}j \ ,
\ee
where $\mathcal{L}_0[f]$ is the classical action for the field $f$, $G[f]$ is the gauge-fixing functional and $\bar{c}QG[f]$ is the ghost Lagrangian. For reducible gauge symmetries there will be additional ghost-for-ghost terms. For a review of the BRST procedure, see \cite{Kugo:1979gm,Henneaux:1992ig, Gomis:1994he,Zoccali:2018pty}. The left and right (tilde) Yang-Mills fields and ghosts transform as:
\be
\label{lin_Yang-Mills_transf}
\begin{split}
Q A_\mu&=\partial_\mu c,\quad Qc=0,\quad Q\bar{c}=\frac{1}{\xi}G(A),\\
Q \tilde{A}_\mu&=\partial_\mu \tilde{c},\quad Q\tilde{c}=0,\quad Q\bar{\tilde{c}}=\frac{1}{\tilde{\xi}}G(\tilde{A}),
\end{split}
\ee
while those of of linearised $\N=0$ supergravity transform as:
\be
\label{flat_grav_BRST}
\begin{array}{lllllllll}
Qh_{\mu\nu}&=&2\partial_{(\mu}c_{\nu)},\quad& Qc_\mu&=&0, \quad& Q\bar{c}_\mu&=&\tfrac{1}{\xi^{(h)}}G_\mu^{(h)},  \\[5pt]
QB_{\mu\nu}&=&2\partial_{[\mu}d_{\nu]},\quad& Qd_\mu&=&\partial_\mu d, \quad&  Q\bar{d}_\mu&=&\tfrac{1}{\xi^{(B)}}G_\mu^{(B)}, \\[5pt]
Q\varphi&=&0.\\
\end{array}
\ee
The BRST system for the two-form additionally contains the (anti)ghost-for-ghosts $\bar{d},d$ and the ghost number 0 object $\eta$, transforming as
\be
Q d=0,\quad Q\bar{d}=\tfrac{1}{\xi_{(d)}}\partial^\mu\bar{d}_\mu,\quad Q\eta=\tfrac{m_{(d)}}{\xi_{(d)}}\partial^\mu d_\mu
\ee
with $\xi_{(d)}$ and $m_{(d)}$ some a priori arbitrary constants. It is convenient to make a choice of gauge fixing functional on the  Yang-Mills  side, and set
\be
\label{flat_gauge_choice}
G[A]\equiv\partial^\mu A_\mu,\quad G[\tilde{A}]\equiv\partial^\mu \tilde{A}_\mu.
\ee 
As derived in \cite{Anastasiou:2018rdx, Borsten:2019prq} a simple dictionary for the linearised fields of $\N=0$ supergravity compatible with the symmetries above is given by
\be
\label{flat_simple_dict}
\begin{aligned}
h_{\mu\nu}=&A_\mu\circ\tilde{A}_\nu+A_\nu\circ\tilde{A}_\mu+a\eta_{\mu\nu}\left(A^\rho\circ\tilde{A}_\rho + \tilde{\xi}c\circ\tilde{\bar{c}}-\xi\bar{c}\circ\tilde{c}\right),\\
B_{\mu\nu}=&A_\mu\circ\tilde{A}_\nu-A_\nu\circ\tilde{A}_\mu, \\
\varphi=&A^\rho\circ\tilde{A}_\rho + \tilde{\xi}c\circ\tilde{\bar{c}}-\xi\bar{c}\circ\tilde{c},
\end{aligned} 
\ee
where $a$ is an arbitrary parameter. 
We can immediately read off the graviton and two-form ghost dictionaries,
\be
\label{lin_gh_dict}
\begin{aligned}
c_\mu=&c\circ\tilde{A_\mu}+A_\mu\circ\tilde{c},\\
d_\mu=&c\circ\tilde{A_\mu}-A_\mu\circ\tilde{c},
\end{aligned} 
\ee
from which the antighost dictionaries follow:
\be
\label{anti_gh_dict}
\begin{aligned}
\bar{c}_\mu=&\bar{c}\circ\tilde{A_\mu}+A_\mu\circ\tilde{\bar{c}},\\
\bar{d}_\mu=&\bar{c}\circ\tilde{A_\mu}-A_\mu\circ\tilde{\bar{c}}.
\end{aligned} 
\ee
Finally, the second order ghosts in the Kalb-Ramond sector are given by
\be 
\label{gh_for_gh_dict}
d=-2c\circ\tilde{c},\quad \bar{d}=-2\bar{c}\circ\tilde{\bar{c}},\quad \eta=-\left(\tilde{\xi}c\circ\tilde{\bar{c}}+\xi\bar{c}\circ\tilde{c} \right)
\ee

A significant advantage of the BRST set-up is that we can directly derive the gauge-fixing functional for the graviton and the two-form, given the Yang-Mills gauge-fixing functional. Indeed, using \eqref{anti_gh_dict} and \eqref{flat_grav_BRST}, in conjunction with the Yang-Mills transformations \eqref{lin_Yang-Mills_transf}, we can determine the graviton and two-form gauge-fixing functionals through 
\be
Q \bar{c}_\mu = \frac{1}{\xi^{\sst{(h)}}} G_\mu^{(h)}, \quad Q \bar{d}_\mu = \frac{1}{\xi^{\sst{(B)}}} G_\mu^{(B)}
\ee
and, making use of the Yang-Mills transformations \eqref{lin_Yang-Mills_transf}, we get
\be 
\begin{aligned}
G_\mu^{(h)}=&\tfrac{\xi^{(h)}\left(\tilde{\xi}+\xi \right)}{2\xi\tilde{\xi}}\left[\partial^\nu h_{\nu\mu}-\tfrac{1}{2}\partial_\mu h + \tfrac{2+(D-2)a}{2}\partial_\mu \varphi\right]+\tfrac{\xi^{(h)}\left(\tilde{\xi}-\xi \right)}{2\xi\tilde{\xi}}\left[\partial^\nu B_{\nu\mu}+\partial_\mu\eta\right],\\
G_\mu^{(B)}=&\tfrac{\xi^{(B)}\left(\tilde{\xi}+\xi \right)}{2\xi\tilde{\xi}}\left[\partial^\nu B_{\nu\mu}+\partial_\mu\eta\right]
+\tfrac{\xi^{(B)}\left(\tilde{\xi}-\xi \right)}{2\xi\tilde{\xi}}\left[\partial^\nu h_{\nu\mu}-\tfrac{1}{2}\partial_\mu h + \tfrac{2+(D-2)a}{2}\partial_\mu \varphi\right].
\end{aligned}
\ee
Note, one can repackage these gauge condition into left/right transverse gauges for the trace-reversed generalised metric $\bar{Z}_{\mu\nu} = \bar{h}_{\mu\nu} + B_{\mu\nu}$,
\be 
\begin{aligned}
G_\mu^{(h)}=&\tfrac{1}{2\xi}\left(\partial^\rho\bar{Z}_{\rho\mu}+\partial_\mu \chi^+ \right)+\tfrac{1}{2\tilde{\xi}}\left(\partial^\rho\bar{Z}_{\mu\rho}+\partial_\mu \chi^{-} \right),\\
G_\mu^{(B)}=&\tfrac{1}{2\xi}\left(\partial^\rho\bar{Z}_{\rho\mu}+\partial_\mu \chi^+ \right)-\tfrac{1}{2\tilde{\xi}}\left(\partial^\rho\bar{Z}_{\mu\rho}+\partial_\mu \chi^{-} \right),
\end{aligned}
\ee
where 
\be
\chi^\pm = \tfrac{(2+Da)}{2}\varphi \pm \eta.
\ee
Here we see the r\^ole of the dilaton appearing in the gauge-fixing functional in direct analogy to the familiar appearance of $\eta$ in the KR gauge-fixing functional. This reflects the fact that it receives  a contribution from the ghost-antighost sector of Yang-Mills squared, $\tilde{\xi}c\circ\tilde{\bar{c}}-\xi\bar{c}\circ\tilde{c}$.

For $\xi=\tilde{\xi}=\xi^{(h)}=\xi^{(B)}$ and $a=\tfrac{2}{2-D}$,  the gauge fixing functionals reduce to
\be 
\label{flat_grav_gauge_simple}
\begin{aligned}
G_\mu^{(h)}=&\partial^\nu h_{\nu\mu}-\tfrac{1}{2}\partial_\mu h, \\
G_\mu^{(B)}=&\partial^\nu B_{\nu\mu}+\partial_\mu\eta,
\end{aligned}
\ee
the natural choices for Einstein frame.

Knowledge of the gauge fixing functionals \eqref{flat_grav_gauge_simple} now allows us to write the linearised Lagrangians:
\be 
\label{lin_BRST_actions}
\begin{aligned}
\hspace{-0.25cm}\mathcal{L}(h,\varphi)=&\tfrac{1}{4}h^{\mu\nu}\square h_{\mu\nu}+\tfrac{1+\xi}{2\xi}\left(\partial^\mu h_{\mu\rho} \partial_\nu h^{\nu\rho}-\partial^\mu h \partial^\nu h_{\mu\nu}\right)-\tfrac{1+2\xi}{8\xi}h\square h-\bar{c}^\mu \square c_\mu +\tfrac{1}{2}\varphi \square \varphi, \\
\mathcal{L}(B)=&\tfrac{1}{4}B^{\mu\nu}\square B_{\mu\nu}+\tfrac{1+\xi}{2\xi}\partial^\mu B_{\mu\nu}\partial_\rho B^{\rho\nu}-\bar{d}^\mu\square d_\mu +\tfrac{\xi}{2}\bar{d}\square d -\tfrac{1}{2\xi}\eta\square\eta.
\end{aligned}
\ee

\subsection{Pure gravity}\label{Killing the dilaton and two-form to get pure gravity}
Here we remove the dilaton and KR two-form to leave pure Einstein-Hilbert gravity. First, it is straightforward to see from \eqref{flat_simple_dict}, \eqref{lin_gh_dict}, \eqref{anti_gh_dict} and \eqref{gh_for_gh_dict} that we can truncate out the entire Kalb-Ramond sector by identifying the two Yang-Mills theories\footnote{Remember that $A_\mu \circ   \tilde{A}_\nu = A_\mu^a \cdot \Phi_{a\tilde{a}} \cdot \tilde{A}_\nu^{\tilde{a}}$, where we are summing over the adjoint indices $a,\tilde{a}$.
In principle, this would allow us to more generally set the two-form sector to vanish without picking $(A_\mu,c,\bar{c})=\alpha(\tilde{A}_\mu,c,\bar{c})$, with $\alpha$ some constant. However, we find it convenient to make this choice, and set $\alpha=1$. }:
\be
A_\mu=\tilde{A}_\mu,\quad c=\tilde{c},\quad \bar{c}=\tilde{\bar{c}}  .
\ee 
To illustrate how the dilaton can be removed, we couple arbitrary sources to the right hand side of the Yang-Mills eom:
\be\label{Yang-Mills_lin_eom}
\square A_\mu-\tfrac{\xi+1}{\xi}\partial_\mu\partial A=j_\mu,\quad \square c=j,\quad \square\bar{c}=\bar{j}.
\ee
Note that, in contrast with the standard treatment of BRST, we have coupled sources to the ghost/antighost. The graviton/dilaton equations, as coming from \eqref{lin_BRST_actions} coupled to sources, become:
\be
\begin{aligned}
\tfrac{1}{2}\square h_{\mu\nu}-\tfrac{1+\xi}{\xi}\partial^\rho\partial_{(\mu}h_{\nu)\rho}+\tfrac{1+\xi}{2\xi}\partial_\mu\partial_\nu h +\eta_{\mu\nu}\left[ \tfrac{1+\xi}{2\xi}\partial^\rho\partial^\sigma h_{\rho\sigma}-\tfrac{1+2\xi}{4\xi}\square h  \right]=&j_{\mu\nu}^{(h)}\\
\square \varphi=&j^{(\varphi)}
\end{aligned}
\ee
Then using \eqref{Yang-Mills_lin_eom} and \eqref{flat_simple_dict}, we can read off the source dictionaries,
\be 
\begin{aligned}
j_{\mu\nu}^{(h)}=&\tfrac{1}{\square}j_\mu\circ j_\nu
-\tfrac{2(1+\xi)}{\square^2}\partial_\mu \partial_\nu j\circ\bar{j}-\tfrac{(\xi+1)^2}{\square^3}\partial_\mu\partial_\nu \partial j\circ \partial j\\
&+\eta_{\mu\nu}\left[\tfrac{\xi(1+\xi)}{\square^2}\partial j\circ\partial j +\tfrac{1+2\xi}{\square}j\circ\bar{j} \right]\\
j^{(\varphi)}=&\tfrac{1}{\square}j^\rho\circ j_\rho+\tfrac{\xi^2-1}{\square^2}\partial j\circ\partial j + \tfrac{2\xi}{\square}j\circ\bar{j},
\end{aligned}
\ee
where from here-on-in we  set $D=4$ for notational clarity, although all of the conclusions hold for arbitrary dimension. 
If we wish to eliminate the dilaton, we first set its source to vanish by picking sources for the ghosts such that
\be
j\circ\bar{j}=-\tfrac{1}{2\xi}j^\rho\circ j_\rho -\tfrac{\xi^2-1}{2\xi}\tfrac{1}{\square}\partial j\circ \partial j
\ee
which allows us to set 
\be\label{dilaton0}
c\circ\bar{c}=-\tfrac{1}{2\xi}A^\rho\circ A_\rho\quad \Rightarrow\quad \varphi=0.
\ee
The graviton source reduces to
\be\label{grav-source}
\begin{aligned}
j_{\mu\nu}^{(0)}=&\tfrac{1}{\square}j_\mu\circ j_\nu
+\tfrac{1+\xi}{\xi \square^2}\partial_\mu\partial_\nu j^\rho\circ j_\rho-\tfrac{(\xi+1)^2}{\xi\square^3}\partial_\mu\partial_\nu \partial j\circ\partial j\\
&~~~-\eta_{\mu\nu}\left[\tfrac{1+2\xi}{2\xi\square}j^\rho\circ j_\rho -\tfrac{(\xi+1)^2}{2\xi\square^2}\partial j \circ \partial j \right]
\end{aligned} 
\ee
and we note that it is unconstrained, even after eliminating the dilaton. On the other hand, in the absence of the ghost contributions setting the dilaton and its source to vanish would constrain the trace of \eqref{grav-source} to be vanishing, cf.~\eqref{source_issue}.  

\noindent Finally, we can invert the dictionaries \eqref{flat_simple_dict}, \eqref{lin_gh_dict} and \eqref{anti_gh_dict} to get:
\be
\label{inverse_dict}
\begin{aligned}
A_\mu\circ A_\nu=&\tfrac{1}{2}h_{\mu\nu}\\
c\circ A_\mu=&\tfrac{1}{2}c_\mu\\
\bar{c}\circ A_\mu=&\tfrac{1}{2}\bar{c}_\mu\\
c\circ\bar{c}=&{-}\tfrac{1}{4\xi}h
\end{aligned}
\ee
and the gauge-fixing functional reduces to the familiar de Donder gauge,
\be
\label{de_donder}
G_\mu[h]=\partial^\nu h_{\nu\mu}-\tfrac{1}{2}\partial_\mu h,
\ee
with the pure gravity BRST action
\be 
\label{lin_BRST_grav}
\mathcal{L}(h)=\tfrac{1}{4}h^{\mu\nu}\square h_{\mu\nu}+\tfrac{1+\xi}{2\xi}\left(\partial^\mu h_{\mu\rho} \partial_\nu h^{\nu\rho}-\partial^\mu h \partial^\nu h_{\mu\nu}\right)-\tfrac{1+2\xi}{8\xi}h\square h-\bar{c}^\mu \square c_\mu.
\ee

\section{Einstein-Hilbert from the double-copy to cubic order}\label{Einstein-Hilbert from the double-copy to cubic order}

\subsection{Gravity as it comes}
We work with the standard Yang-Mills BRST action
\be 
\mathcal{L}_{\text{YM}}=-\tfrac{1}{4}F_{\mu\nu}^a F^{\mu\nu a}+\tfrac{1}{2\xi}G[A]^aG[A]^a-\bar{c}^a\partial^\mu D_\mu^{ac}c^c,
\ee
where $F_{\mu\nu}^a=\partial_\mu A_\nu^a-\partial_\nu A_\mu^a+gf^{abc}A_\mu^bA_\nu^c$, $D_\mu^{ac}=\delta^{ac}\partial_\mu+gf^{abc}A_\mu^b$ and the gauge-fixing functional is linear $G[A]^{a}=\partial^\rho A^{a}_{\rho}$. Note, here Feynman gauge corresponds to $\xi=-1$. Up to cubic order this becomes
\be 
\begin{aligned}
\mathcal{L}_{\text{YM}}=&-\tfrac{1}{4}F_{\mu\nu}^{a(0)}F^{\mu\nu a(0)}+\tfrac{1}{2\xi}\partial^\rho A^{a}_{\rho}\partial^\rho A^{a}_{\rho}-\bar{c}^a\square c^a\\
&-gf^{abc}\partial_\mu A_\nu^a A^{\mu b}A^{\nu c}-gf^{abc}\bar{c}^a\partial^\mu\left(A_\mu^b c^c \right),
\end{aligned}
\ee
with $F^{(0)}=\partial_\mu A_\nu^a-\partial_\nu A_\mu^a$. The cubic terms can be written as 
\be
\begin{split}
\mathcal{L}^{(3)}_{\text{YM}}&=igf^{abc}\int \dbar p_1 \dbar p_2 \dbar p_3\ e^{-i(p_1+p_2+p_3)x} \left[ \tfrac{1}{6}n^{\mu\nu\rho}(p_i) A_\mu^a(p_1)A_\nu^b(p_2)A_\rho^c(p_3) \right.\\
&~~\phantom{=\tfrac{ig}{6}f^{abc}\int \dbar p_1 \dbar p_2 \dbar p_3\ e^{-i(p_1+p_2+p_3)x}} \left.+n^{\mu\alpha\beta}(p_1){c}^{a}_{\alpha}(p_1)A_\mu^b(p_2)c^{c}_{\beta}(p_3) \right],
\end{split}
\ee
with $\dbar p=\tfrac{d^4p}{\left(2\pi \right)^4}$ and we have isolated the BCJ satisfying, in the sense that it is totally antisymmetric, kinematic numerator
\be
n^{\mu_1\mu_2\mu_3}(p_i)=-(p_{12}^{\mu_3}\eta^{\mu_1\mu_2}+p_{23}^{\mu_1}\eta^{\mu_2\mu_3}+p_{31}^{\mu_2}\eta^{\mu_3\mu_1}),
\ee
where $p_{ij}=p_i-p_j$.
The above is as in classical Yang-Mills\footnote{This is a consequence of choosing a linear gauge-fixing functional. It would be interesting to study models where $G$ is a nonlinear function of $A_\mu$.}, however note that we now have a contribution coming from the ghost-antighost-gluon interaction term with kinetic numerator 
\be
n^{\mu\alpha\beta}(p)=-p^\mu\sigma_{+}^{\alpha\beta},\qquad \sigma_{\pm}=\tfrac{1}{2}\left(\sigma_x\pm i\sigma_y\right),
\ee
where for convenience we have introduced the ghost-antighost doublet, $c_\alpha = (c, \bar{c})$. Here $\sigma_i$ are the Pauli matrices and $\sigma_{+}^{\alpha\beta}c_\alpha c_\beta$ creates a ghost number zero state.

When performing the double-copy, we must take all possible combinations: the gluon-gluon-gluon term with itself will contribute graviton-graviton-graviton interactions, same as the ghost-antighost-gluon term with itself, while the cross terms will contribute the graviton-ghost-antighost interactions.  Note, we have no \emph{a priori} reason to believe the ghost sector of gravity generated through the na\"ive double-copy presented above will be consistent. However, as we shall demonstrate,  it yields up to field redefinitions  Einstein-Hilbert gravity with BRST respecting gauge-fixing and ghost sectors, as required. 

To implement the BCJ double-copy we introduce a super-index $M = (\mu, \alpha)$ and send $if^{abc}X^{{M}{N}{P}}\rightarrow \alpha_{\sst{({MNP}, \tilde{M}\tilde{N}\tilde{P})}}X^{\tilde{M}\tilde{N}\tilde{P}}X^{{M}{N}{P}}$, where there is no sum between the set of normalisation parameters,  $\alpha$, and the non-zero components of $X$, which are determined by the allowed diagrams,
\be
X^{\mu\nu\rho}=n^{\mu\nu\rho},\qquad X^{\mu\alpha\beta}=n^{\mu\alpha\beta},\qquad X=0~~\text{otherwise}.
\ee
This yields the momentum space double-copy Lagrangian,
\be 
\begin{aligned}
\hat{\mathcal{L}}^{(3,dc)}_{(grav)}&=\alpha_{\sst{({MNP}, \tilde{M}\tilde{N}\tilde{P})}}X^{\tilde{M}\tilde{N}\tilde{P}}  \left[ \tfrac{1}{6}n^{\mu\nu\rho}A_{\mu\tilde{M}}A_{\nu\tilde{N}}A_{\rho\tilde{P}} +n^{\nu\alpha\beta}{c}_{\alpha\tilde{M}}A_{\nu\tilde{N}}c_{\beta\tilde{P}} \right] \\
&= \tfrac{1}{6}\alpha_1  n^{\mu\nu\rho} n^{\tilde{\mu}\tilde{\nu}\tilde{\rho}} A_{\mu\tilde{\mu}}(p) A_{\nu\tilde{\nu}}(k) A_{\rho\tilde{\rho}}(q) + \alpha_3 p^\mu p^{\tilde{\mu}}\bar{C}^{(0)}(p)A_{\mu\tilde{\mu}}(k)C^{(0)}(q) \\
&~~~-2\alpha_2^+ n^{\mu\nu\rho}p^{\tilde{\nu}} \bar{C}_\mu(p)A_{\nu\tilde{\nu}}(k)C_\rho(q) ,
\end{aligned}
\ee
where for convenience we have labelled the non-zero constants \footnote{Note that only a linear combination of the normalisation parameters  $\alpha_{\sst{({\mu\nu\rho}, \tilde{\mu}\tilde{\alpha}\tilde{\beta})}}$   and $\alpha_{\sst{({\mu\alpha\beta}, \tilde{\mu}\tilde{\nu}\tilde{\rho})}}$ will be fixed through the double copy. This is a consequence of the fact that we have restricted to the symmetric sector in order to focus on pure gravity. In the full construction, the orthogonal combination
\be
 \alpha_2^-=\tfrac{1}{2}\left( \tfrac{1}{6}\alpha_{\sst{({\mu\nu\rho}, \tilde{\mu}\tilde{\alpha}\tilde{\beta})}}- \alpha_{\sst{({\mu\alpha\beta}, \tilde{\mu}\tilde{\nu}\tilde{\rho})}}\right)
\ee
will be fixed by studying the ghost sector of the two-form $B_{\mu\nu}$ \cite{Nagy2020}.} 
\be
\alpha_1=\alpha_{\sst{({\mu\nu\rho}, \tilde{\mu}\tilde{\nu}\tilde{\rho})}} ,\qquad \alpha_2^+=\tfrac{1}{2}\left( \tfrac{1}{6}\alpha_{\sst{({\mu\nu\rho}, \tilde{\mu}\tilde{\alpha}\tilde{\beta})}}+ \alpha_{\sst{({\mu\alpha\beta}, \tilde{\mu}\tilde{\nu}\tilde{\rho})}}\right) , \qquad \alpha_3= \alpha_{\sst{({\mu\alpha\beta}, \tilde{\mu}\tilde{\alpha}\tilde{\beta})}}
\ee
and 
defined
\be 
\begin{aligned}
A_{\mu\nu}=&\mathcal{F}[A_\mu\circ A_\nu]\\
C_\mu=&\mathcal{F}[A_\mu\circ c],\quad \bar{C}_\mu=\mathcal{F}[A_\mu\circ \bar{c}] \\
C^{(0)}=&\mathcal{F}[c\circ \bar{c}]=-\bar{C}^{(0)}
\end{aligned}
\ee
with $\mathcal{F}$ denoting the  Fourier transform. Making use of the linear dictionary \eqref{inverse_dict}, the graviton sector of  the above can be rewritten in position space as
\be 
\label{cubic_BCJ_action}
\begin{aligned}
\mathcal{L}^{(3,dc)}_{(grav)}=&-\tfrac{\alpha_1}{8}h^{\mu\nu}\left(h^{\rho\sigma}\partial_\rho\partial_\sigma h_{\mu\nu}- \partial_\mu h^{\rho\sigma}\partial_\nu h_{\rho\sigma}- h^{\rho\sigma}\partial_\sigma\partial_\nu h_{\mu\rho} \right.\\
&\left.~~~~~~~~~~~+2\partial_\nu h_{\rho\sigma}\partial^\sigma h_\mu^{\ \rho}- \partial_\rho h_{\nu\sigma}\partial^\sigma h_\mu^{\ \rho}-\tfrac{ \alpha_3}{4\xi^2\alpha_1}h\partial_\mu\partial_\nu h\right).
\end{aligned}
\ee 
Note that this is simpler than the Einstein-Hilbert action at cubic order given in \eqref{class_cubic_action}, thus revealing one of the advantages of the double copy dictionary. As observed in \cite{Bern:2010yg}, the double-copy of the purely gluonic sector of the Yang-Mills  action performed in this manner will give a graviton action that correctly reproduces the on-shell amplitudes to this order. The terms of \eqref{class_cubic_action} that vanish in the on-shell amplitude, due to the transverse-traceless polarisation tensors and momentum conservation, simply do not appear here.  However, the ghost-antighost sector reintroduces a term depending on $h$, which allows one to fix  the Einstein-frame dilaton to vanish, as will be demonstrated in \autoref{match}.  Said another way, the vanishing of  the Einstein-frame dilaton at linear order given in \eqref{dilaton0} remains  consistent at higher orders. 

\subsection{Matching to perturbative Einstein-Hilbert gravity}\label{match}

At cubic level, the standard BRST action for gravity is
\be 
\label{cubic_BRST_action}
\mathcal{L}^{3,BRST}=\mathcal{L}_{class}^{(3)}+\tfrac{1}{\xi}G_\mu[h]^{(1)}G^\mu[h]^{(2)}-\left\{\bar{c}^\mu Q\left[G_\mu[h]\right] \right\}^{(3)},
\ee
with the superscripts denoting the order in perturbation theory. $\mathcal{L}_{class}^{(3)}$ is just the cubic part of the Einstein-Hilbert action:
\be
\label{class_cubic_action}
\begin{split} 
\mathcal{L}_{class}^{(3)}=
&
\frac{1}{2}h^{\mu \nu } \left(\tfrac{1}{2}  \partial_{\mu }h^{\rho \sigma} \partial_{\nu }h_{\rho \sigma}-  \tfrac{1}{4}\eta_{\mu\nu} \partial_{\sigma}h_{\tau \rho} \partial^{\sigma}h^{\tau \rho}+  \partial_{\nu }h\left( \partial_{\rho }h_{\mu }{}^{\rho } -  \tfrac{1}{2}  \partial_{\mu }h \right) \right.
\\
&
+ \partial_{\nu }h_{\mu }{}^{\rho } \partial_{\rho }h - \partial_{\rho }h \partial^{\rho }h_{\mu \nu } -  \tfrac{1}{2}\eta_{\mu\nu} \partial^{\rho }h \left(\partial_{\sigma}h_{\rho }{}^{\sigma} - \tfrac{1}{2} \partial_{\rho }h  \right)
+  \partial^{\rho }h_{\mu \nu } \partial_{\sigma}h_{\rho }{}^{\sigma}  
\\
&
- 2  \partial_{\nu }h_{\rho \sigma} \partial^{\sigma}h_{\mu }{}^{\rho }
 -  \partial_{\rho }h_{\nu \sigma} \partial^{\sigma}h_{\mu }{}^{\rho } + \partial_{\sigma}h_{\nu \rho} \partial^{\sigma}h_{\mu }{}^{\rho } 
\left. + \tfrac{1}{2}\eta_{\mu\nu} \partial_{\rho }h_{\tau \sigma} \partial^{\sigma}h^{\tau \rho} \right).
\end{split}
\ee
The linear part of the gauge fixing functional is determined via the BRST procedure in \eqref{de_donder}:
\be
G_\mu[h]^{(1)}=\left[\partial^\nu h_{\nu\mu}-\tfrac{1}{2}\partial_\mu h \right] ,
\ee
while the second order part $G_\mu[h]^{(2)}$ is to be determined by matching with the BCJ action \eqref{cubic_BCJ_action}. The normalisation factors in the double copy dictionary are fixed to
\be 
\alpha_1=1,\quad \alpha_3=\xi^2,
\ee
and we find that one needs to perform a nonlinear field redefinition of the graviton fluctuation
\be \label{redef}
h_{\mu\nu}\to h_{\mu\nu} - \tfrac{1}{4}   h_{\mu\nu}h +  \tfrac{1}{2}   h_{\mu}^{\ \rho}h_{\nu\rho}
  -\tfrac{1}{16}   \eta_{\mu\nu} \left( h_{\rho\sigma}h^{\rho\sigma}   -\tfrac{3}{4}   h^2\right).
\ee
Comparing \eqref{cubic_BRST_action} with \eqref{cubic_BCJ_action}, we derive the next order in the gauge fixing functional prior to the field redefintion
\be 
\begin{split}
G_\mu[h]^{(2)}&=  \tfrac{3}{8}   h^{\nu\rho}\partial_\mu h_{\nu\rho}  -\tfrac{5+2\xi}{32}   h\partial_\mu h+  \tfrac{4-3\xi}{16}   h_\mu^{\ \rho}\partial_\rho h   \\
&~~~~ -\tfrac{1}{2}   h^{\nu\rho}\partial_\rho h_{\mu\nu}+  \tfrac{1}{4}  h\partial^\rho h_{\mu\rho}  -\tfrac{4-\xi}{4}   h_\mu^{\ \nu}\partial^\rho h_{\nu\rho}.
\end{split}
\ee
Once the field redefinition \eqref{redef} is applied, the gauge-fixing functional simplifies and is proportional to the free parameter $\xi$, as expected,
\be\label{gf}
G_\mu[h]^{(2)}\to  \tfrac{\xi}{8}\left(  h_{\mu}{}^{\nu}  \partial_{\rho}  h_{\nu}{}^{\rho} -\tfrac{1}{2}    h \partial_{\mu}  h -  \tfrac{3}{2}    h_{\mu}{}^{\nu} \partial_{\nu}  h   \right).
\ee 
Note, restricting to local field redefinitions, \eqref{redef} and \eqref{gf} are uniquely determined. Note moreover, there is no local field redefinition matching Einstein-Hilbert without the ghost-antighost sector\footnote{Recall, we are setting the dilaton in Einstein-frame to vanish, and  thus,  forbidding further field redefinitions.}.

Once the gauge fixing term at second order has been found, the ghost terms in the cubic action are uniquely determined by the last term in \eqref{cubic_BRST_action}, together with the perturbative BRST transformation
\be 
Q h_{\mu\nu}=2\partial_{(\mu}c_{\nu)}+\kappa \left[c^\rho\partial_\rho h_{\mu\nu}-2 c^\rho\partial_{(\mu}h_{\nu)\rho}\right]
\ee 
This can be matched to the BCJ ghost terms
\be 
\begin{aligned}
\mathcal{L}^{(3,dc)}_{(gh)}
=&-\tfrac{\alpha_2^+}{4} \left(  h_{\nu\rho}\partial^\rho\bar{c}_\mu\partial^\nu c^\mu-   h_{\mu\rho} \partial^\rho\bar{c}_\nu\partial^\nu c^\mu-   \partial_\mu h_{\nu\rho}\partial^\rho\bar{c}^\nu c^\mu\right.\\
&~~~~\left.+   \partial_\nu h_{\mu\rho} \partial^\rho\bar{c}^\nu c^\mu+    h_{\nu\rho}\partial^\rho\partial_\mu \bar{c}^\nu c^\mu-   h_{\nu\rho}\partial^\rho\partial^\nu\bar{c}_\mu c^\mu\right)
\end{aligned}
\ee
by fixing the normalisation parameter
\be 
 \alpha_2^+=1
\ee
and performing a  non-local transformation on the ghost and antighost fields, which is not unique. A convenient, in the sense that it places no restrictions on the range of $\xi$, example is given by:
\be \label{credef}
\begin{aligned}
c_\mu&\to c_\mu+  \tfrac{3\xi-4}{32} h c_\mu +\tfrac{3}{4} h_\mu^{\ \nu}c_\nu+\tfrac{1}{2\square}\Big[\tfrac{\xi+3}{8}\partial_\rho h \partial_\mu c^\rho
 +\tfrac{1}{2}\partial_\mu h\partial_\rho c^\rho\\&
~~~ -\tfrac{1}{2} \partial^\sigma h_{\mu\sigma}\partial^\rho c_\rho -\tfrac{\xi+2}{4}\partial^\rho h_{\sigma\rho}\partial_\mu c^\sigma-\tfrac{\xi+2}{4}\partial^\rho h_{\sigma\rho}\partial^\sigma c_\mu
+\tfrac{\xi+1}{2} h_{\mu\rho}\partial^\rho\partial^\sigma c_\sigma\Big]\\
\bar{c}_\mu&\to\bar{c}_\mu-\tfrac{3\xi+4}{32} h\bar{c}_\mu -\tfrac{\xi+2}{8} h_\mu^{\ \nu}\bar{c}_\nu +\tfrac{1}{\square}\Big[-\tfrac{2\xi^2-3\xi+4}{16\xi} \partial_\rho\partial_\mu h\bar{c}^\rho -\tfrac{\xi-1}{4\xi} \partial_\mu\partial^\sigma h_{\sigma\rho}\bar{c}^\rho\\&~~~+\tfrac{1}{4\xi}\partial_\sigma\partial_\rho h_\mu^{\ \sigma} \bar{c}^\rho
-\tfrac{1+2\xi}{4\xi} \partial^\rho\partial^\sigma h_{\rho\sigma}\bar{c}_\mu -\tfrac{3\xi^2-12\xi-4}{32\xi}\square h\bar{c}_\mu
+\tfrac{1+2\xi}{16} h\partial_\mu\partial_\rho \bar{c}^\rho \Big]\\
h_{\mu\nu}&\to h_{\mu\nu}-\tfrac{1}{2} \bar{c}_{(\mu}c_{\nu)}
\end{aligned}
\ee 
Although the above is non-local, of course, the resulting action is local. This follows from the fact that the linear component of the gauge-fixing functional was determined by the double-copy to be de Donder \eqref{de_donder}. This yields a specific form for the quadratic ghost action \eqref{lin_BRST_grav} proportional to $\bar{c}_\rho \Box c^\rho$, prior to any field redefinitions,  which excludes all possibly non-local terms that may have arisen from \eqref{credef}.

\section{Conclusions} \label{Conclusions}

In this paper, we demonstrated that the BRST convolution product, in conjunction with the BCJ algorithm, can reproduce the Lagrangian of pure BRST Einstein-Hilbert gravity up to cubic order. We found that the ghost sector of the Yang-Mills  action played a crucial r\^ole in achieving this in the pure graviton sector. We additionally derived the gauge fixing functional up to second order in fluctuations and the corresponding diffeomorphism ghost action.

We have focused here on the pure gravity case as a proof of principle, however the full $\N=0$ supergravity construction, including the two-form and dilaton, would be of interest, both conceptually and from the perspective of classical solutions \cite{Goldberger:2016iau,Goldberger:2017ogt,Luna:2016hge,Kim:2019jwm}. Work on this is in progress \cite{Nagy2020}.  

Another obvious generalisation would be to promote one of the factors to a full off-shell $\N=1$ vector supermultiplet as in \cite{Anastasiou:2014qba}. In this case, the ghost-antoghost sector would produce an entire chiral multiplet, reflecting the fact that on-shell ``$\N=0$ Yang-Mills $\times$ $\N=1$ Yang-Mills'' yields $\N=1$ supergravity coupled to a chiral multiplet.  

We also note that we have made a choice of a linear gauge fixing functional $G[A]^{a}=\partial^\rho A^{a}_{\rho}$ for the YM theory. It would very instructive to study non-linear gauge choices - the challenge in this context would be to understand how the BCJ rules need to be modified. 

Note, we have from the beginning eliminated the Nakanishi-Lautrup auxiliary field corresponding to the Yang-Mills antighost. It would perhaps be instructive to understand what r\^ole it might play in the convolution product. The  full Batalin-Vilkovisky formalism and BRST complex will be treated, from an independent perspective not relying on the convolution product, in work in progress \cite{Borsten:2020}. 

Of course, an important question is how to proceed to higher orders in perturbation theory. A path towards this would possibly make use of the BCJ respecting Yang-Mills Lagrangians of \cite{Bern:2010yg} and \cite{Tolotti:2013caa}, which include identically vanishing non-local terms to all orders that then need to be made local and cubic through the introduction of auxiliary fields as described at four and five points in \cite{Bern:2010yg}.

 \acknowledgments

We are grateful to Leonardo de la Cruz, Donal O'Connell, Andres Luna, Ricardo Monteiro and Chris White for helpful discussions. LB is particularly grateful to Hyungrok Kim for illuminating conversations. SN is supported by STFC grant ST/P000703/1 and a Leverhulme Research Project Grant. The work of LB has been supported by a Schr\"odinger Fellowship and the Leverhulme Trust.

\bibliography{Ref_Library}
\bibliographystyle{utphys}


\end{document}